\documentclass[aps,prl,twocolumn,citeautoscript,superscriptaddress]{revtex4}
\usepackage{amsmath}
\usepackage{amssymb}
\usepackage{graphicx,bm}
\usepackage[caption=false]{subfig}
\usepackage{verbatim}
\usepackage{epstopdf}
\usepackage{dcolumn}
\usepackage{floatrow}
\usepackage{float}
\usepackage{booktabs}
\usepackage[utf8]{inputenc}
\usepackage{array}
\usepackage{makecell}
\usepackage{booktabs}
\usepackage{multirow}
\usepackage{dcolumn}
\usepackage{footnote}

\usepackage{lipsum}

\usepackage{hyperref}
\hypersetup{
	colorlinks=true,
	linkcolor=magenta,
	filecolor=violet,     
	urlcolor=blue,
	citecolor=red
}

\begin{document}
	
\title{Fragmentation dynamics of diatomic molecules under proton impact: Kinetic energy release spectra of $CO^{q+}$ and $NO^{q+}$ ($q = 2, 3$) molecular ions.}

\author{Avijit Duley}
\affiliation{Indian Institute of Technology Kanpur, Kanpur - 208016, India}

\author{Narendra Nath Dutta}
\affiliation{School of Sciences, SR University, Warangal}

\author{C. Bagdia}
\affiliation{Tata Institute of Fundamental Research, 1 Homi Bhabha Road, Colaba, Mumbai - 400005, India}

\author{L. C. Tribedi}
\affiliation{Tata Institute of Fundamental Research, 1 Homi Bhabha Road, Colaba, Mumbai - 400005, India}

\author{C. P. Safvan}
\affiliation{Inter-Univerosty Accelerator Centre, Aruna Asif Ali Marg, New Delhi - 110067, India}

\author{A. H. Kelkar}
\email{akelkar@iitk.ac.in}
\affiliation{Indian Institute of Technology Kanpur, Kanpur - 208016, India}

\begin{abstract}
We report on the fragmentation dynamics of triply charged, diatomic, molecular ions of $NO$ and $CO$. Dissociative fragmentation after multiple ionization of $NO$ and $CO$ is studied under the impact of 200 keV proton beam using recoil-ion momentum spectrometer. Kinetic Energy Release distributions (KERDs) for various fragmentation channels were obtained. We have also calculated the potential energy curves (PECs) for ground and several excited states of $NO^{3+}$ and $CO^{3+}$ molecular ion. The obtained KERDs are discussed in the background of the calculated PECs as well as the simple Coulomb excitation model. Coulomb breakup of the unstable precursor molecular ion shows a clear preference for the $N^{2+} + O^+$ (and $C^{2+} + O^+$) fragmentation channel.
\end{abstract}
\maketitle

\section{I. INTRODUCTION}
The study of ionization and fragmentation dynamics of small and large molecules is an active area of atomic physics research. Fragmentation of molecular ions leads to creation of atomic ion and free radicals. The creation and evolution of such ionic and neutral fragments is a key aspect in understanding the processes related to plasma physics \cite{Plasma}, atmospheric and space physics \cite{RevModPhys.85.1021}, radiation damage \cite{PhysRevLett.91.053401,PhysRevLett.107.023202}etc. In a laboratory environment the precursor molecular ion can be generated by collisions of neutral molecules with photons, electrons or heavy ion projectile beams. The advent of recoil ion momentum spectrometer (RIMS) \cite{Ullrich_1997,Ullrich_2003}, along with position sensitive detector allows for detailed investigation of the subsequent molecular fragmentation process. In the past couple of decades, there have been extensive studies on the dissociation dynamics of multiply charged molecular ions produced in collisions with heavy ions \cite{Ullrich_1997,Ullrich_2003,PhysRevA.47.3748,PhysRevA.57.2608}, electrons \cite{Fainelli1996Mar,Pandey1,Pandey2,PhysRevA.87.022709} and photons \cite{Curtis1985Feb,PhysRevLett.82.2075}. Although, the target molecules under investigation range from diatomic molecules %\cite{PhysRevA.74.032701,Khan_2021} 
to polyatomic systems %\cite{Kelkar_2007,PhysRevA.75.041201,PhysRevA.82.043201,PhysRevA.90.032701} 
including large bio-molecules %\cite{PhysRevA.85.032711,PhysRevLett.91.053401,Agnihotri_2013} 
and PAHs (poly-cyclic aromatic hydrocarbons) %\cite{PhysRevA.104.L060802}, 
it is the diatomics which have been studied extensively.

Several studies (experimental and theoretical) have investigated the ionization and fragmentation of simple diatomics such as $N_2$, $O_2$ and $CO$ under particle %\cite{PhysRevA.62.022718,PhysRevA.82.054702,Werner_2001,Melo_2008,Folkerts1996Oct,Sharma_2019,Sharma_2018,PhysRevA.97.020701} 
and photon impact %\cite{0953-4075-25-6-011,PhysRevLett.93.113003,PhysRevA.95.060702,doi:10.1063/1.3436722,doi:10.1063/1.5130706,PhysRevA.48.4379,PhysRevA.54.2004,PhysRevA.92.013402,PhysRevA.100.013415}. 
In collisions where the ionization and subsequent Coulomb fragmentation of the precursor molecular ion take place at time scales shorter than the vibrational time period of the molecular ion, the kinetic energy release distribution (KERD) provides complete information of the internal excitation of the molecular ion. The KER values observed experimentally can be directly compared with calculated potential energy curves (PECs) for the neutral and ionized molecule. The electronic states and corresponding PECs for neutral, singly ionized and doubly ionized diatoms are readily available in literature \cite{Tobias,O'Neil,TMasuoka,Hurley,BibEntry2020Jul}, however, such calculations for multiply charged molecular ions are rather limited \cite{VKrishnamurthi,MLarsson,LCederbaum}. For example, there are very few reports on the fragmentation dynamics of $NO$ molecule \cite{PWThulstrup,PhysRevA.100.053422,Erman_1996,Lai_2016} and the PEC calculations are available only up to doubly charged ($NO^{2+}$) molecular ions \cite{BibEntry2020Jul,PWThulstrup,DLCooper}.

In order to produce multiply charged molecular ions, sufficient energy needs to be deposited in the neutral molecule to cause multi-electron ionization/excitation. Collisions with energetic heavy ion beams are the most efficient way of producing multiple ionization in the parent molecule. Additionally, in collisions with heavy ion beams with hundreds of keV energy, the interaction time is $\le$ femtoseconds. This interaction time is shorter than the typical vibrational and rotational time scales of the molecules and the electronic excitation of the molecule is a Frank-Condon transition. Therefore, the KER depends on the internuclear separation which is characters tic of the precursor molecular ion state. 

In this paper, We present our combined experimental and theoretical study on the fragmentation of triply charged $NO^{3+}$ and $CO^{3+}$ molecular ions produced in collisions of the neutral molecules with 250 keV proton beam. Kinetic energy release distributions have been obtained for $NO^{q+}$ and $CO^{q+}$ ($q = 2,3$) molecular ions. We have also calculated the energy of excited electronic states and corresponding PECs for triply charged molecular ions. The theoretical PECs for $CO^{3+}$ are compare well with existing calculations, whereas for $NO^{3+}$ molecular ion the PECs have not been reported previously.  

\begin{figure}
\raggedright
\includegraphics[width=\textwidth]{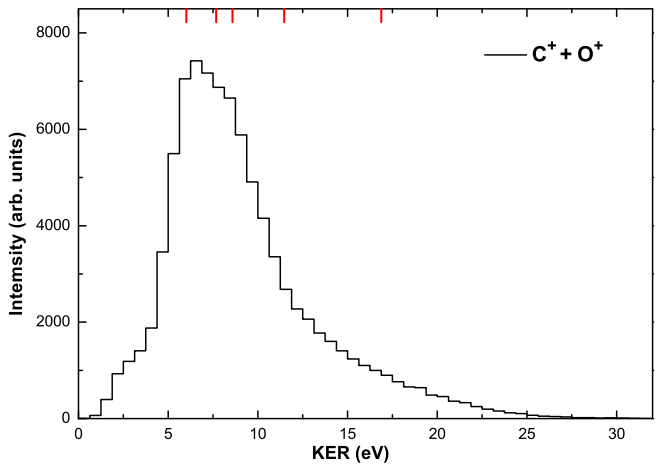}
\caption{Kinetic energy release spectra for $CO^{2+}$ fragmentation}
\label{Fig.1}
\end{figure}

\begin{figure}
\raggedright
\includegraphics[width=\textwidth]{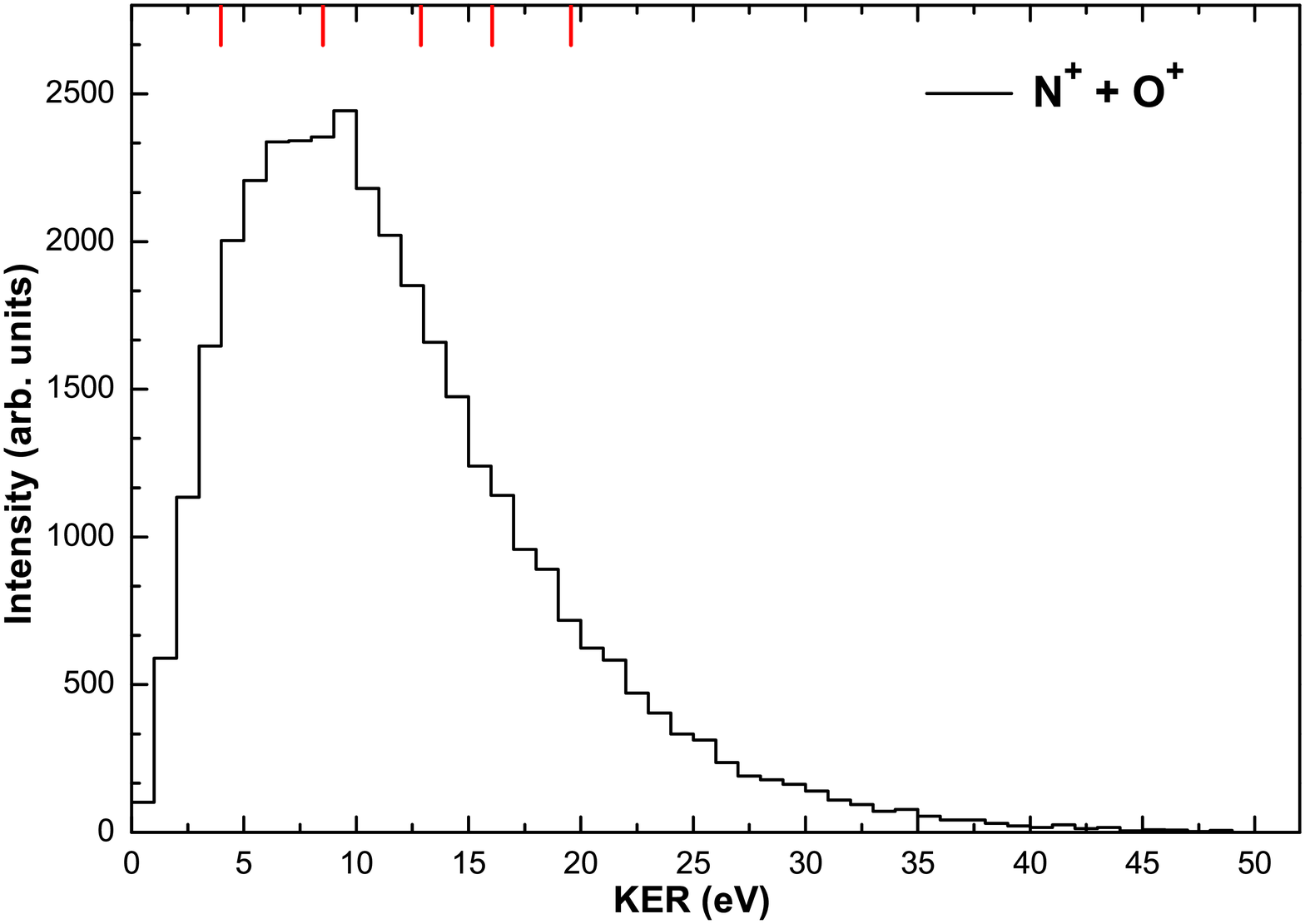}
\caption{Kinetic energy release spectra for $NO^{2+}$ fragmentation}
\label{Fig.2}
\end{figure}

\begin{table}
\begin{tabular*}{\textwidth}{l  @{\extracolsep{\fill}}  c c r }
\hline\hline \\ 
{Fragmentation} & KER     & Coulomb                                       & Relative  \\ 
       	{Channel}       & $(eV)$  & energies                                      & Intensity \\ 
       	                      &         & $(eV)$             & $(\%)$    \\ [0.25ex]\\
\hline \\ 
                                    	$O^{+}+O^{+}$       & $9.5$    & $11.9$   & $93.38$ \\
                                    	$O^{2+}+O^{+}$      & $19.4$   & $23.8$   & $6.62$  \\  
	                                	$C^{+}+O^{+}$       & $6 \pm 0.6$    & $12.8$ \footnote{\label{note1}Ref\cite{PhysRevA.75.062709}}    & $91.51$ \\ 
	                                	$C^{2+}+O^{+}$      & $16\pm1$   & $25.5$\textsuperscript{\ref{note1}}    & $6.76$  \\ 
	                                	$C^{+}+O^{2+}$      & $19\pm1$   & $25.5$\textsuperscript{\ref{note1}}    & $1.73$  \\  
	                                	$N^{+}+O^{+}$       & $8.52\pm2$   & $12.52$    & $69.1$ \\ 
	                                	$N^{2+}+O^{+}$      & $29\pm2$   & $25.04$    & $20.97$  \\ 
	                                	$N^{+}+O^{2+}$      & $32\pm2$   & $25.04$    & $9.93$  \\  [0.25ex]\\ 
	                                	 
		\hline \hline
	\end{tabular*}
	\caption{Observed most probable values in the KER spectra for the various fragmentation channels for $CO^{q+}$ and $NO^{q+}$ ($q = 2, 3$) fragmentation induced by proton impact.The values calculated from a pure Coulomb explosion model are also included.}
	\label{Table I}
\end{table}

\section{II. EXPERIMENT AND DATA ANALYSIS}

The experiments were performed at two accelerator facilities in India. Measurements with $CO$ molecular target were performed at 300 keV ECR ion accelerator (ECRIA) at TIFR, Mumbai, and for $NO$ molecular target, the experiments were performed at the low energy ion beam facility (LEIBF) at IUAC, New Delhi. The time-of-flight mass spectrometer (TOFMS) setup at both the accelerator facilities are very similar and have been described in detail elsewhere \cite{Shubhadeep,LEIBF2}. Briefly, a collimated beam of 200 keV (250 keV at IUAC) protons was made to collide with an effusive molecular gas target in a high vacuum scattering chamber recoil ions, following collision, were extracted by a TOFMS located perpendicular to the projectile ion beam as well as the gas jet. A channel electron multiplier was used to detect the electrons emitted in the ionization process and signal was used as start signal for the data acquisition system. The dissociated ionic fragments were detected on the other end of the TOFMS using a micro-channel plate (MCP) detector equipped with a position sensitive delay line anode (DLD). Signals from the MCP and the DLD were fed to a multi-hit time to digital converter (tdc). The tdc allows detection of both ionic fragments from a single dissociation event. Analysis of the coincidence data was performed on a event by event basis to obtain the 3D momenta of the ionic fragments. The three momenta of each ion pair were then used to obtain the kinetic energy release distributions (KERD) for a given dissociation channel.

In the present projectile energy range, charge exchange processes such as electron capture may also contribute towards multiple ionization of the diatomic molecules. However, in the present experiments we have not analyzed the projectile beam post collisions and the data presented are integrated over all processes leading to multiple ionization and fragmentation of the parent molecule.

\section{III. THEORETICAL POTENTIAL ENERGY CURVES}
In order to calculate the potential energy curves (PECs) for the two ions CO$^{3+}$ and NO$^{3+}$, we calculate the PECs for the ground states of CO and NO molecules firstly. The ground states of CO and NO molecules are $^1\Sigma^{+}$ and $^2\Pi$, respectively. The PECs for the $^1\Sigma^{+}$ and $^2\Pi$ states are calculated using multireference configuration interaction (MRCI) method with complete active space self-consistent field (CASSCF) reference functions. For CASSCF calculations, we consider the full-valence type active space. We use the MOLPRO software package \cite{MOLPRO} which uses internally contracted version of MRCI approach (icMRCI) \cite{doi:10.1063/1.455556, KNOWLES1988514, doi:10.1063/1.3609809} for robust and accurate generations of the PECs at various internuclear separations (grid points). The basis set used in our calculations is correlation consistent cc-pV5Z basis set.

\begin{figure}[t]
\raggedright
\includegraphics[width=\textwidth]{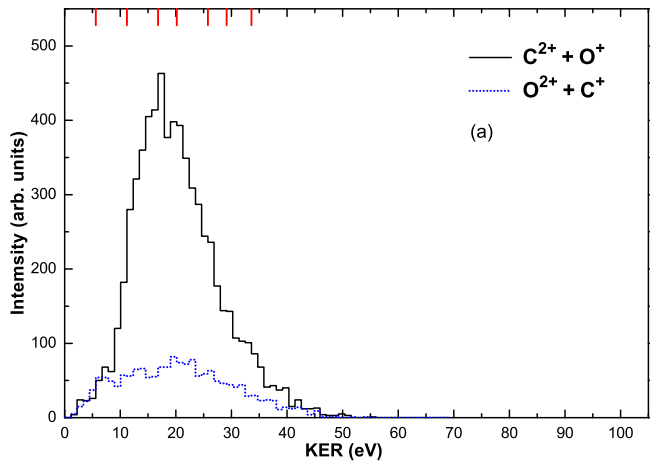}
\caption{Kinetic energy release spectra for $CO^{3+}$ fragmentation}
\label{Fig.3}
\end{figure}

We calculate these PECs for sufficiently large set of grid points starting from 0.8 $\AA$ to 4.0 $\AA$ to ensure the position of the minimum for each of the both molecules CO and NO. The minimums for the states $^1\Sigma^{+}$ and $^2\Pi$ for CO and NO, respectively, occur at internuclear separations of 1.13 $\AA$ and 1.15 $\AA$, respectively, with dissociation energies of 11.17 eV and 6.46 eV, respectively. The corresponding experimental minimum is reported as 1.128 $\AA$ \cite{cccbdb} and dissociation energy is estimated around 11.09 eV \cite{Huber1979} for CO. Whereas the experimental values are 1.154 $\AA$ \cite{cccbdb} and 6.496 eV \cite{Huber1979}, respectively, for NO. The PECs for the two molecular ions CO$^{3+}$ and NO$^{3+}$ are calculated with respect to the total molecular energies obtained at the minimums of these $^1\Sigma^{+}$ and $^2\Pi$ states, respectively. We adopt a very much similar strategy to calculate the PECs for these ionic systems as we consider for their neutral counterparts, i.e., icMRCI approach based on CASSCF reference functions with full-valence active space and cc-pV5Z basis sets. However, unlike CO and NO, our aim is to calculate 3 states for each of the 8 symmetries $^2\Pi$, $^4\Pi$, $^2\Delta$, $^4\Delta$, $^2\Sigma^{+}$, $^4\Sigma^{+}$, $^2\Sigma^{-}$ and $^4\Sigma^{-}$ for CO$^{3+}$, and  $^1\Pi$, $^3\Pi$, $^1\Delta$, $^3\Delta$, $^1\Sigma^{+}$, $^3\Sigma^{+}$, $^1\Sigma^{-}$ and $^3\Sigma^{-}$ for NO$^{3+}$. This is the reason we choose state-average CASSCF approach for each symmetries to have a balanced descriptions of the orbitals for all the 3 states of a particular symmetry. The grid points for the PECs of these ionized species are calculated from 0.9 $\AA$ to 2.00 $\AA$ with grid separation of 0.05 $\AA$ and then from 2.00 $\AA$ to 3.00 $\AA$ with grid separation of 0.10 $\AA$. However, the grid points are adjusted (by minimal amount) in a very few cases where proper convergence is not achieved during icMRCI calculations. In general, calculations of the PECs up to 3 $\AA$ is sufficient to identify the channels of atomic fragmentation for the 24 states calculated for each of the triply ionized molecules. From around 2.5 $\AA$, the PECs are almost consistent with the fall with increasing internuclear separation as expected following the equation $E_{AB}=E_A+E_B+\frac{28.8}{R^2}$ \cite{Pandey}. Here $E_{AB}$ is the total energy of the molecular state, $E_A$ and $E_B$ are the energies of the fragmented ions at the dissociation limit of the state, and $R$ is the internuclear separation in $\AA$. The last term of this equation represents the Coulomb repulsion energy (in eV) between a singly and a doubly ionized atomic species at a distance $R$ apart. The vertical excitation energies for the first states of $^4\Sigma^{-}$, $^2\Sigma^{+}$, $^2\Pi$ and $^2\Delta$ symmetries of CO$^{3+}$ are calculated to be 81.92 eV, 83.17 eV, 81.32 eV, and 83.83 eV, respectively. These results agree quite well with similar calculations with cc-pVQZ basis sets performed by Kumar and Sathyamurthy \cite{5}. They obtained these values 81.82 eV, 82.63 eV, 81.26 eV, and 83.13 eV, respectively. Also, our estimated value of 83.83 eV for first state of $^2\Delta$ symmetry is in good agreement with the experimental value of 83.4 eV \cite{Mathur, WATSON1975384}. The energy states of the fragmented ions, C$^{+}$, C$^{++}$, O$^{+}$ and O$^{++}$, are calculated using the same icMRCI approach and same quality of basis functions. The excitation energies between the different states are compared with the corresponding values obtained from the atomic spectra database of National Institute of Standards and Technology (NIST) \cite{NIST_ASD} and good agreements are found between these two sets of energies.

\begin{figure}[t]
\raggedright
\includegraphics[width=\textwidth]{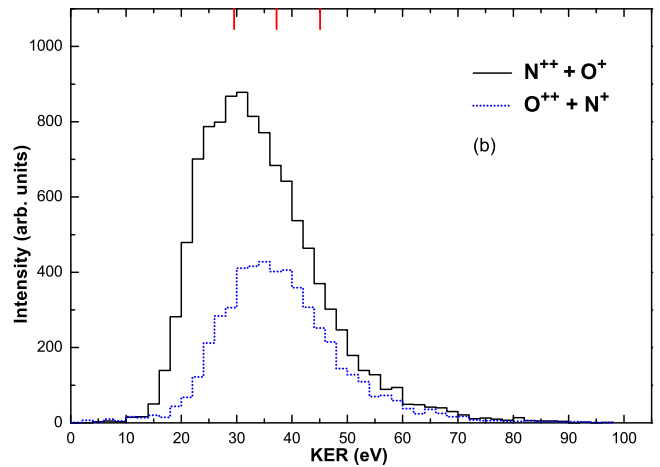}
\caption{Kinetic energy release spectra for $NO^{3+}$ fragmentation}
\label{Fig.4}
\end{figure}

\section{IV. RESULTS AND DISCUSSIONS}

In ion-molecule collision experiments, the molecular ion can be created in multiple charge states depending on the the degree of parent ionization. The multiply ionized parent molecular ion is very short lived due to mutual Coulomb repulsion between the two (for diatomic molecules) nuclear centers and dissociates in to fragment ions. In previous studies with highly charged heavy ions with $CO$, multiply charged molecular ions up to $CO^{7+}$ have been observed \cite{PhysRevA.75.062709}. On the other hand few hundred keV protons act as relatively soft projectiles resulting in mainly doubly and triply ionized molecular ions. In the present measurements we have observed parent molecular ions up to $CO^{4+}$ and $NO^{4+}$, however the discussion is limited to the dissociation of doubly and triply charged molecular ions as the yield for quadruply charged ion was rather small. We have measured the kinetic energy release distributions for a) the single symmetric dissociation channel of $CO^{2+}$ and $NO^{2+}$ ions and b) the two asymmetric dissociation channels of $CO^{3+}$ and $NO^{3+}$ molecular ions. In table \ref{Table I} we have listed the most probably KER values for each dissociation channel. The KER values calculated using the simple Coulomb explosion model are also shown in the same table. It is evident that, in general, the most probable KER value is smaller than that predicted by the Coulomb explosion model. The Coulomb explosion model is a simple model describing the fragmentation of molecular ions. In this model, the individual atomic centers of the parent molecular ion are considered as point charges and the Coulomb repulsion between these atoms determines the kinetic energy release of the dissociation process. The energy expected from this model is given by: $E(eV)=14.4 \frac{q_{1}q_{2}}{R_{e}(\AA)}$, where $q_1$ and $q_2$ are the asymptotic charges of the two fragments, and $R_{e}$ is the equilibrium internal nuclear distance of the neutral molecule. $R_{e}$ corresponds to the distance from which vertical transition is expected to take place according to the Frank-Condon principle. However, the model excludes the effect of the charge cloud in the inter nuclear region as well as the electron correlation effects. Therefore, the Coulomb explosion model is known to predict KER values higher than those obtained experimentally  \cite{PhysRevA.74.032701}. 

\subsection{B. Dissociation of $CO^{2+}$ and $NO^{2+}$}

The experimentally observed KER spectra for dissociation of doubly charged $CO^{2+}$ and $NO^{2+}$ molecular ions are shown in Fig.\ref{Fig.1} and Fig.\ref{Fig.2}. The doubly charged parent molecular ion dissociates into singly charged fragments, $C^+ + O^+$ (and $N^+ + O^+$) due to mutual repulsion, which is also the most dominant fragmentation channel for this collision system. Dissociation into a charged and a neutral fragment can not be observed with this experimental setup. The KER spectrum for $CO^{2+}$ fragmentation channel after collision with photons \cite{TMasuoka1,Xiaokai} and charged particles \cite{Pandey1,Pandey2,Mathur,PhysRevA.75.062709,Watson1996Feb} has been reported by several groups. The most probable KER value measured in this experiment (see Table\ref{Table I}) and the overall shape of the KER distribution is in agreement with the reported values in literature \cite{PhysRevA.75.062709,Mathur,Watson1996Feb,Wells2005Aug,Ben-Itzhak1995Jan,Folkerts1996Oct}.

\begin{figure}
\raggedright
\includegraphics[width=8.5cm]{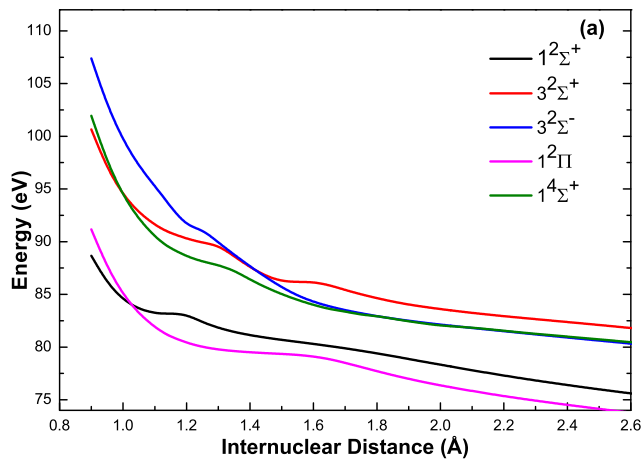}
\caption{Potential Energy Curves from CASSCF-MRCI calculations on  $CO^{3+}$ }
\label{Fig.5}
\end{figure}

\begin{figure}
\includegraphics[width=8.5cm]{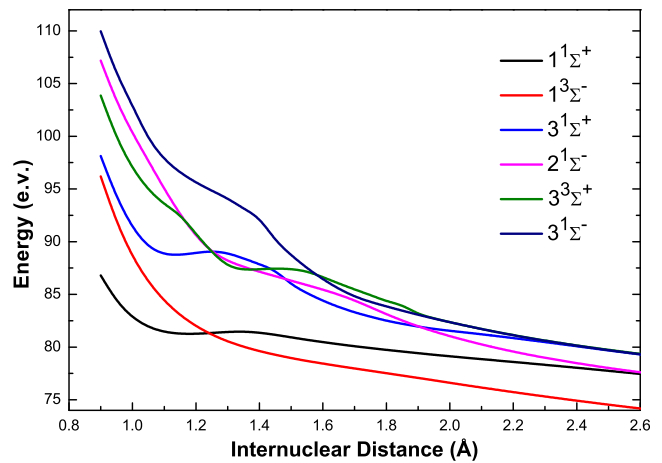}
\caption{Potential Energy Curves from CASSCF-MRCI calculations on  $NO^{3+}$}
\label{Fig.6}
\end{figure}
 
The observed KER values can be accounted for by considering three low lying electronic states of $CO^{2+}$ ion \cite{PhysRevLett.75.1058}. The KER spectra contains contributions from $1^{1}\Sigma^{+}$,$1^{1}\Pi$ and $X^{3}\Pi$ dissociating into ground state products of $C^{+}(^{2}P)+O^{+}(^{4}S)$. It is to be noted that the KERD has a long tail up to 30 eV which may arise due to the participation of higher electronic states of the precursor ion. Dissociation dynamics of the $NO$ molecule has not been studied as extensively as the $CO$ molecular target. Nevertheless, few experimental studies have been reported for $NO^{2+}$ fragmentation in collisions with photons \cite{Curtis1984Oct,Curtis1985Feb,Masuoka1994May} and electrons \cite{Fainelli1996Mar,YBKim}. The KER spectrum is similar to that obtained for $CO^{2+}$ ion. The potential energy curves (PECs) for $NO^{2+}$ have been calculated by various groups \cite{PWThulstrup,DLCooper,Hurley}. One can associate the measured KER spectra with a few prominent PECs. For both, $CO^{2+}$ and $NO^{2+}$ fragmentation, it is seen that the most probable KER value is much smaller than that calculated using the Coulomb explosion model. However, the tail of KER spectra extends well beyond this value on the high energy side. Such large KER can be attributed to the high lying electronic states of the doubly charged parent molecular ion. The high lying electronic states are more repulsive due to a steeper slope in the Frank-Condon region owing to weaker screening of the positive nuclear centers. This results in a higher KER of the fragment ions. 

\subsection{C. Dissociation of $CO^{3+}$ and $NO^{3+}$}

The dissociation of a triply charged hetero atomic molecular ion such as $CO^{3+}$ and $NO^{3+}$ can proceed via two channels where the total electronic charge of the parent molecular ion is shared asymmetrically by the two fragment ions.  The two possible fragmentation channels are represented as channel \ref{ch1} and channel \ref{ch2} as below:
\begin{table}[H]
	\begin{tabular*}{0.9\textwidth}{@{} c @{\extracolsep{\fill}}c c  c @{}}
		\hline\hline \\ [0.25ex]
	
		KER        &  KER                     & Molecular   & Dissociation  \\ 
	    $(eV)$     &  $(eV)$                  & states      &limit\\
		(calc.)   &\cite{Mathur}                   &             &     \\[0.25ex]
	
		\hline  \\ [0.25ex]
		
		14.42 &13.74         & $2^{2}\Sigma^{+}$      &$C^{2+}(^{3}P)+O^{+}(^{2}D)$   \\
		
		16.97 & -      & $1^{2}\Sigma^{+}$ & $C^{2+}(^{1}S)+O^{+}(^{2}P)$  \\ 
		
		 17.21 & 17.85      & $1^{4}\Pi$       & $C^{2+}(^{3}P)+O^{+}(^{4}S)$  \\

		 17.76 & 17.53      & $2^{2}\Sigma^{+}$       & $C^{2+}(^{3}P)+O^{+}(^{4}S)$  \\
		    
		   18.13 & -      & $1^{2}\Pi$       & $C^{2+}(^{1}S)+O^{+}(^{2}D)$  \\
		
		19.26 & -      & $2^{2}\Sigma^{+}$       & $C^{2+}(^{1}S)+O^{+}(^{2}P)$  \\  
		
		     19.30 & -      & $2^{4}\Pi$       & $C^{2+}(^{3}P)+O^{+}(^{2}D)$  \\  
		     
		     19.53 &19.55      & $1^{2}\Sigma^{-}$       & $C^{2+}(^{1}S)+O^{+}(^{2}D)$  \\ 
		     
		     19.83 & -      & $2^{2}\Pi$       & $C^{2+}(^{1}S)+O^{+}(^{2}P)$  \\ 
		     
		     19.93 & -      & $3^{2}\Sigma^{+}$       & $C^{2+}(^{3}P)+O^{+}(^{2}D)$  \\ 
		     
		    20.06 & -      & $1^{4}\Sigma^{+}$       & $C^{2+}(^{3}P)+O^{+}(^{2}D)$  \\ 
		     
		    20.49 & -      & $3^{2}\Pi$       & $C^{2+}(^{3}P)+O^{+}(^{4}S)$  \\ 
		     
		     20.64 & 21.25      & $1^{2}\Delta$       & $C^{2+}(^{1}S)+O^{+}(^{2}D)$  \\
		     
		     21.21 & -      & $1^{4}\Delta$       & $C^{2+}(^{3}P)+O^{+}(^{2}D)$  \\
		     
		     22.85 & -      & $2^{4}\Sigma^{+}$       & $C^{2+}(^{3}P)+O^{+}(^{2}P)$  \\
		     
		     22.89 & -      & $2^{2}\Delta$       & $C^{2+}(^{3}P)+O^{+}(^{2}D)$  \\
		     
		     22.94 & -      & $3^{4}\Sigma^{-}$       & $C^{2+}(^{3}P)+O^{+}(^{2}D)$  \\
		     
		     23.44 & -      & $3^{2}\Delta$       & $C^{2+}(^{3}P)+O^{+}(^{2}D)$  \\
		     
		     23.52 & -      & $2^{4}\Sigma^{-}$       & $C^{2+}(^{3}P)+O^{+}(^{4}S)$  \\

		24.29  & -      & $3^{4}\Pi$       & $C^{2+}(^{3}P)+O^{+}(^{2}D)$  \\ 
		
		     24.37  & -      & $2^{4}\Delta$       & $C^{2+}(^{3}P)+O^{+}(^{2}D)$  \\ 
		     
		     24.44  & -      & $3^{2}\Delta$       & $C^{2+}(^{3}P)+O^{+}(^{2}D)$  \\ 
		     
		     24.42 & -      & $3^{2}\Sigma^{-}$       & $C^{2+}(^{3}P)+O^{+}(^{2}D)$  \\
		     
		     26.37 & -      & $2^{2}\Sigma^{-}$       & $C^{2+}(^{3}P)+O^{+}(^{4}S)$  \\[0.05ex] \\    
		\hline \\ [0.05ex]

		19.07 & -      & $3^{2}\Sigma^{+}$       & $C^{+}(^{2}P)+O^{2+}(^{3}P)$  \\
		   
		    23.66 & -      & $2^{4}\Sigma^{+}$       & $C^{+}(^{2}P)+O^{2+}(^{3}P)$  \\[0.05ex] \\
	
		\hline \hline
	\end{tabular*}
	\caption{The possible molecular states of $CO^{3+}$ dissociating into $C^{2+}+O^+$ and $C^{+}+O^{2+}$ along with the KER peaks obtained by fitting the experimental KERD, theoretically calculated values of KER. The KER values are also compared with that from Ref.\cite{Mathur}}
	\label{Table II}
\end{table}
\begin{equation}
    \begin{split}
    CO^{3+} \longrightarrow C^{2+} + O^{+}\\
    NO^{3+} \longrightarrow N^{2+} + O^{+}
\end{split}
\label{ch1}
\end{equation}
\begin{equation}
    \begin{split}
    CO^{3+} \longrightarrow C^{+} + O^{2+}\\
    NO^{3+} \longrightarrow N^{+} + O^{2+}
\end{split}
\label{ch2}
\end{equation}
The KER spectra for these two fragmentation channels are shown in the Fig.\ref{Fig.3} and Fig.\ref{Fig.4}. It is observed that channel \ref{ch1} is the dominant mode for charge asymmetric dissociation. In Table \ref{Table I}, we have shown the relative intensities of channel \ref{ch1} and channel \ref{ch2}. In the case of $CO^{3+}$ dissociation, the measured yield of channel \ref{ch1} is $\sim$ 4 times larger than channel \ref{ch2}. The same in the case of $NO^{3+}$ dissociation is found to be $\sim$ 2. The dominance of channel \ref{ch1} over channel \ref{ch2} has also been seen in previous studies with $CO$ molecular target \cite{0953-4075-30-24-019, PhysRevA.47.2827, Wells2005Aug, PhysRevA.46.3929}. The reported ratio varies from 2 to 5 depending on the charge, mass, and energy of the projectile ion. Similar data for $NO^{3+}$ dissociation, however has not been reported so far. The difference in the yield of the two dissociation channels can be explained on the basis of ionization potential of the two participating atoms. The ionization potential of Carbon (and Nitrogen) atom is smaller compared to that of Oxygen atom. Subsequently, the total energy needed to create the final dissociation state is 49.3 eV and 64.2 eV ($CO^{3+}$ and $NO^{3+}$ respectively) for channel \ref{ch1} as opposed to 59.9 eV and 69.7 eV for channel \ref{ch2} \cite{Mathur, Suzuki1995}. Relatively larger yield of channel \ref{ch2} in the case of $NO^{3+}$ dissociation is in agreement with the smaller difference in ionization energies of the final state fragment ions.

The KER spectra for the two channel are shown in Fig.\ref{Fig.3} and Fig.\ref{Fig.4}. The most probable values of KERD differ slightly as listed in Table \ref{Table I}. The KER spectrum for $CO^{3+}$ fragmentation can be compared with previous experimental investigations. The most probable KER value depends on the choice of projectile and values ranging from 11 eV to 27 eV have been reported for channel \ref{ch1} \cite{Ben-Itzhak1995Jan,PhysRevA.87.022709,PhysRevA.75.062709}. For channel \ref{ch2} the most probable KER value is generally higher than that for channel \ref{ch1}. The difference in most probable KER appears to be strongly dependent on the choice of projectile. For example, with hevay ion projectiles in the MeV range, the most probable KER for dissociation via channel \ref{ch2} has been shown to be $\sim$ 10 eV larger than that for channel \ref{ch1}. In collisions with photons, this difference is much smaller \cite{GWei_2009}. In the present experiments with 200 keV protons, this difference is also not very large (see table \ref{Table I}). The KER spectra can be compared with the calculated potential energy curves for $CO^{3+}$ and $NO^{3+}$ molecular ions. As stated earliar, numerous theoretical calculations are available for $CO^{3+}$. However, the theoretical potential energy curves for excited states of $NO^{3+}$ molecular ions have not been reported. In Table \ref{Table II} and \ref{Table III}, we have listed the calculated PECs for $CO^{3+}$ and $NO^{3+}$ molecular ions respectively. The observed KER spectra are in good agreement with the calculated PEC values. In Fig \ref{Fig.5} and \ref{Fig.6} we have also shown the PECs for a few selected states. The steep slope and repulsive nature of excited states is evident leading to higher KER. We also observe that for the case of $NO^{3+}$ fragmentation, the higher excited states appear to contributing dominantly resulting in a larger value of most probable KER value. 

\begin{table}
	\begin{tabular*}{0.9\textwidth}{ c @{\extracolsep{\fill}}c c  @{}}
		\hline\hline \\ [0.25ex]
		KER                         & Molecular  & Dissociation  \\ 
	    $(eV)$                     & states      &limit\\
		(calc.)                   &             &     \\[0.25ex]
		\hline  \\ [0.25ex]
		
		14.42   &$1^{1}\Sigma^{+}$   &  $N^{++}(^{2}P)+O^{+}(^{2}D)$   \\ 
		17.76   &$1^{1}\Sigma^{+}$   &  $N^{++}(^{2}P)+O^{+}(^{4}S)$   \\ 
		15.37   &$1^{1}\Pi$   &  $N^{++}(^{2}P)+O^{+}(^{2}D)$   \\
		16.45   &$2^{1}\Sigma^{+}$   &  $N^{++}(^{2}P)+O^{+}(^{2}P)$   \\
	16.72   &$1^{3}\Sigma^{+}$   &  $N^{++}(^{2}P)+O^{+}(^{2}D)$   \\
		17.56   &$1^{1}\Pi$   &  $N^{++}(^{2}P)+O^{+}(^{4}S)$   \\
		17.58   &$1^{1}\Delta$   &  $N^{++}(^{2}P)+O^{+}(^{2}D)$   \\
		18.94   &$2^{3}\Pi$   &  $N^{++}(^{2}P)+O^{+}(^{2}D)$   \\
		19.71   &$2^{3}\Sigma^{+}$   &  $N^{++}(^{2}P)+O^{+}(^{4}S)$   \\
		20.22   &$3^{1}\Sigma^{+}$   &  $N^{++}(^{2}P)+O^{+}(^{2}P)$   \\
   	    21.89   &$3^{1}\Sigma^{+}$   & $N^{++}(^{2}P)+O^{+}(^{2}D)$   \\
		21.23   &$3^{3}\Pi$   &  $N^{++}(^{2}P)+O^{+}(^{2}D)$   \\
		21.81   &$2^{1}\Pi$   &  $N^{++}(^{2}P)+O^{+}(^{2}D)$   \\
		22.13   &$1^{3}\Delta$   &  $N^{++}(^{2}P)+O^{+}(^{2}D)$   \\
		23.30   &$2^{3}\Sigma^{-}$   &  $N^{++}(^{2}P)+O^{+}(^{2}D)$   \\
		23.32   &$1^{1}\Sigma^{-}$   &  $N^{++}(^{2}P)+O^{+}(^{2}D)$   \\
		23.35   &$3^{3}\Sigma^{+}$   &  $N^{++}(^{2}P)+O^{+}(^{2}D)$   \\
		24.02   &$2^{1}\Delta$   &  $N^{++}(^{2}P)+O^{+}(^{2}D)$   \\
	24.07   &$3^{3}\Sigma^{+}$   &  $N^{++}(^{2}P)+O^{+}(^{2}P)$   \\
		24.49   &$3^{3}\Delta$   &  $N^{++}(^{2}P)+O^{+}(^{2}P)$   \\
		24.70   &$3^{1}\Pi$   &  $N^{++}(^{2}P)+O^{+}(^{2}D)$   \\
		25.72   &$2^{1}\Sigma^{-}$   &  $N^{++}(^{2}P)+O^{+}(^{2}D)$   \\
		26.13   &$3^{3}\Sigma^{-}$   &  $N^{++}(^{2}P)+O^{+}(^{2}D)$   \\
		27.99   &$3^{1}\Delta$   &  $N^{++}(^{2}P)+O^{+}(^{2}P)$   \\
		28.02   &$3^{1}\Sigma^{-}$   &  $N^{++}(^{2}P)+O^{+}(^{2}P)$   \\ [0.05ex] \\
		
		\hline \\ [0.05ex] 
		
		19.69   &$3^{1}\Sigma^{+}$   &  $N^{+}(^{3}P)+O^{++}(^{3}P)$   \\
		23.55   &$3^{3}\Sigma^{+}$   &  $N^{+}(^{3}P)+O^{++}(^{3}P)$   \\
		23.97   &$3^{3}\Delta$   &  $N^{+}(^{3}P)+O^{++}(^{3}P)$   \\
		27.47   &$3^{1}\Delta$   &  $N^{+}(^{3}P)+O^{++}(^{3}P)$   \\
		27.49   &$3^{1}\Sigma^{-}$   &  $N^{+}(^{3}P)+O^{++}(^{3}P)$   \\ [0.05ex] \\

		\hline \hline
	\end{tabular*}
	\caption{The possible molecular states of $NO^{3+}$ dissociating into $N^{2+}+O^+$ and $O^{2+}+N^{+}$ along with the theoretically calculated values of KER}
	\label{Table III}
\end{table}

\section{V. CONCLUSION}
In this work, we have investigated the dissociation dynamics of multiply charged molecular ions of $CO$ and $NO$. The parent molecular ions are created in ionizing collisions with 200 keV protons. The dissociation pathways and corresponding KER spectra were measured using a recoil ion momentum spectrometer. We have also calculated the potential energy curves for the triply charged $CO^{3+}$ and $NO^{3+}$ ions. The measured KER distribution is found match well with the calculated PECs. For the case of $NO^{3+}$ fragmentation, the most probable KER value is found to be slightly larger than the one calculated from a simple Coulomb explosion model. This is attributed to the existence of high lying excited states of $NO^{3+}$ molecular ions.    

\section{ACKNOWLEDGEMENT}
We are very much thankful to Dr. K. R. Shamasundar, Indian Institute of Science Education and Research (IISER) Mohali, India for his valuable advice to compute the PECs and to give us access of the SHREE cluster of IISER Mohali. We would like to acknowledge Dr. Satyam Srivastava, Indian Association for the Cultivation of Science (IACS), Kolkata, India and Dr. Amrendra Pandey, Universit$\acute{\text{e}}$ Paris-Saclay, France for providing us some valuable tipsvaluable suggestions to calculate the PECs.   

\bibliographystyle{utphys}
\bibliography{References}

\providecommand{\href}[2]{#2}\begingroup\raggedright\begin{thebibliography}{10}

\bibitem{Plasma}
R.~K. Janev, ``Atomic and molecular processes in fusion edge plasmas.,'' {\em
  New York: Plenum Press. 1995} .

\bibitem{RevModPhys.85.1021}
A.~G. G.~M. Tielens, ``The molecular universe,''
  \href{http://dx.doi.org/10.1103/RevModPhys.85.1021}{{\em Rev. Mod. Phys.}
  {\bfseries 85} (Jul, 2013) 1021--1081}.
  \url{https://link.aps.org/doi/10.1103/RevModPhys.85.1021}.

\bibitem{PhysRevLett.91.053401}
J.~de~Vries, R.~Hoekstra, R.~Morgenstern, and T.~Schlath\"olter, ``Charge
  driven fragmentation of nucleobases,''
  \href{http://dx.doi.org/10.1103/PhysRevLett.91.053401}{{\em Phys. Rev. Lett.}
  {\bfseries 91} (Jul, 2003) 053401}.
  \url{https://link.aps.org/doi/10.1103/PhysRevLett.91.053401}.

\bibitem{PhysRevLett.107.023202}
P.~L\'opez-Tarifa, M.-A. Herv\'e~du Penhoat, R.~Vuilleumier, M.-P. Gaigeot,
  I.~Tavernelli, A.~Le~Padellec, J.-P. Champeaux, M.~Alcam\'{\i},
  P.~Moretto-Capelle, F.~Mart\'{\i}n, and M.-F. Politis, ``Ultrafast
  nonadiabatic fragmentation dynamics of doubly charged uracil in a gas
  phase,'' \href{http://dx.doi.org/10.1103/PhysRevLett.107.023202}{{\em Phys.
  Rev. Lett.} {\bfseries 107} (Jul, 2011) 023202}.
  \url{https://link.aps.org/doi/10.1103/PhysRevLett.107.023202}.

\bibitem{Ullrich_1997}
J.~Ullrich, R.~Moshammer, R.~Dörner, O.~Jagutzki, V.~Mergel,
  H.~Schmidt-Böcking, and L.~Spielberger, ``Recoil-ion momentum
  spectroscopy,'' \href{http://dx.doi.org/10.1088/0953-4075/30/13/006}{{\em
  Journal of Physics B: Atomic, Molecular and Optical Physics} {\bfseries 30}
  no.~13, (Jul, 1997) 2917--2974}.
  \url{https://doi.org/10.1088%2F0953-4075%2F30%2F13%2F006}.

\bibitem{Ullrich_2003}
J.~Ullrich, R.~Moshammer, A.~Dorn, R.~D. rner, L.~P.~H. Schmidt, and H.~S.-B.
  cking, ``Recoil-ion and electron momentum spectroscopy:
  reaction-microscopes,''
  \href{http://dx.doi.org/10.1088/0034-4885/66/9/203}{{\em Reports on Progress
  in Physics} {\bfseries 66} no.~9, (Aug, 2003) 1463--1545}.
  \url{https://doi.org/10.1088%2F0034-4885%2F66%2F9%2F203}.

\bibitem{PhysRevA.47.3748}
I.~Ben-Itzhak, K.~D. Carnes, S.~G. Ginther, D.~T. Johnson, P.~J. Norris, and
  O.~L. Weaver, ``Fragmentation of ${\mathrm{ch}}_{4}$ caused by fast-proton
  impact,'' \href{http://dx.doi.org/10.1103/PhysRevA.47.3748}{{\em Phys. Rev.
  A} {\bfseries 47} (May, 1993) 3748--3757}.
  \url{https://link.aps.org/doi/10.1103/PhysRevA.47.3748}.

\bibitem{PhysRevA.57.2608}
D.~L. Hansen, M.~E. Arrasate, J.~Cotter, G.~R. Fisher, K.~T. Leung, J.~C.
  Levin, R.~Martin, P.~Neill, R.~C.~C. Perera, I.~A. Sellin, M.~Simon,
  Y.~Uehara, B.~Vanderford, S.~B. Whitfield, and D.~W. Lindle, ``Neutral
  dissociation of hydrogen following photoexcitation of hcl at the chlorine $k$
  edge,'' \href{http://dx.doi.org/10.1103/PhysRevA.57.2608}{{\em Phys. Rev. A}
  {\bfseries 57} (Apr, 1998) 2608--2611}.
  \url{https://link.aps.org/doi/10.1103/PhysRevA.57.2608}.

\bibitem{Fainelli1996Mar}
E.~Fainelli, F.~Maracci, R.~Platania, and L.~Avaldi, ``{Auger
  electron{\textendash}ion coincidence experiment on nitric oxide molecule
  excited by electron impact},'' \href{http://dx.doi.org/10.1063/1.471088}{{\em
  J. Chem. Phys.} {\bfseries 104} no.~9, (Mar, 1996) 3227--3233}.

\bibitem{Pandey1}
S.~K.~R. Pandey~A., Bapat~B., ``Charge symmetric dissociation of doubly ionized
  n2 and co molecules,'' \href{http://dx.doi.org/10.1063/1.4861665}{{\em The
  Journal of Chemical Physics} {\bfseries 140} no.~1, (2014) 034319}.
  \url{https://pubmed.ncbi.nlm.nih.gov/25669391/}.

\bibitem{Pandey2}
A.~Pandey, P.~Kumar, S.~B. Banerjee, K.~P. Subramanian, and B.~Bapat,
  ``Electron-impact dissociative double ionization of ${\mathrm{n}}_{2}$ and
  co: Dependence of transition probability on impact energy,''
  \href{http://dx.doi.org/10.1103/PhysRevA.93.042712}{{\em Phys. Rev. A}
  {\bfseries 93} (Apr, 2016) 042712}.
  \url{https://link.aps.org/doi/10.1103/PhysRevA.93.042712}.

\bibitem{PhysRevA.87.022709}
R.~Singh, P.~Bhatt, N.~Yadav, and R.~Shanker, ``Ionic fragmentation of the co
  molecule by impact of 10-kev electrons: Kinetic-energy-release
  distributions,'' \href{http://dx.doi.org/10.1103/PhysRevA.87.022709}{{\em
  Phys. Rev. A} {\bfseries 87} (Feb, 2013) 022709}.
  \url{https://link.aps.org/doi/10.1103/PhysRevA.87.022709}.

\bibitem{Curtis1985Feb}
D.~M. Curtis and J.~H.~D. Eland, ``{Coincidence studies of doubly charged ions
  formed by 30.4 nm photoionization},''
  \href{http://dx.doi.org/10.1016/0168-1176(85)80028-8}{{\em Int. J. Mass
  Spectrom. Ion Processes} {\bfseries 63} no.~2, (Feb, 1985) 241--264}.

\bibitem{PhysRevLett.82.2075}
P.~Bolognesi, D.~B. Thompson, L.~Avaldi, M.~A. MacDonald, M.~C.~A. Lopes, D.~R.
  Cooper, and G.~C. King, ``Vibrationally selected
  ${O}^{+}\ensuremath{-}{O}^{+}$ fragmentation of ${O}_{2}$ below the adiabatic
  double-ionization potential studied via electron-electron coincidence
  spectroscopy,'' \href{http://dx.doi.org/10.1103/PhysRevLett.82.2075}{{\em
  Phys. Rev. Lett.} {\bfseries 82} (Mar, 1999) 2075--2078}.
  \url{https://link.aps.org/doi/10.1103/PhysRevLett.82.2075}.

\bibitem{Tobias}
I.~Tobias, R.~J. Fallon, and J.~T. Vanderslice, ``Potential energy curves for
  co,'' \href{http://dx.doi.org/10.1063/1.1731475}{{\em The Journal of Chemical
  Physics} {\bfseries 33} no.~6, (1960) 1638--1640},
  \href{http://arxiv.org/abs/https://doi.org/10.1063/1.1731475}{{\ttfamily
  https://doi.org/10.1063/1.1731475}}. \url{https://doi.org/10.1063/1.1731475}.

\bibitem{O'Neil}
S.~V. O'Neil and H.~F. Schaefer, ``Valence‐excited states of carbon
  monoxide,'' \href{http://dx.doi.org/10.1063/1.1673871}{{\em The Journal of
  Chemical Physics} {\bfseries 53} no.~10, (1970) 3994--4004},
  \href{http://arxiv.org/abs/https://doi.org/10.1063/1.1673871}{{\ttfamily
  https://doi.org/10.1063/1.1673871}}. \url{https://doi.org/10.1063/1.1673871}.

\bibitem{TMasuoka}
T.~Masuoka, ``Kinetic‐energy release in the dissociation of co2+,''
  \href{http://dx.doi.org/10.1063/1.468192}{{\em The Journal of Chemical
  Physics} {\bfseries 101} no.~1, (1994) 322--327},
  \href{http://arxiv.org/abs/https://doi.org/10.1063/1.468192}{{\ttfamily
  https://doi.org/10.1063/1.468192}}. \url{https://doi.org/10.1063/1.468192}.

\bibitem{Hurley}
A.~C. Hurley and V.~W. Maslen, ``{Potential Curves for Doubly Positive Diatomic
  Ions},'' \href{http://dx.doi.org/10.1063/1.1731793}{{\em J. Chem. Phys.}
  {\bfseries 34} no.~6, (Jun, 1961) 1919--1925}.

\bibitem{BibEntry2020Jul}
Jul, 2020.
\newblock \url{https://pubs.acs.org/doi/pdf/10.1021/ja00737a003}. [Online;
  accessed 6. Jul. 2020].

\bibitem{VKrishnamurthi}
V.~Krishnamurthi, K.~Nagesha, V.~R. Marathe, and D.~Mathur, ``Probing the
  quantal identity of low-lying electronic states of ${\mathrm{co}}^{2+}$ by
  quantum-chemical calculations and ion-translational-energy spectrometry,''
  \href{http://dx.doi.org/10.1103/PhysRevA.44.5460}{{\em Phys. Rev. A}
  {\bfseries 44} (Nov, 1991) 5460--5467}.
  \url{https://link.aps.org/doi/10.1103/PhysRevA.44.5460}.

\bibitem{MLarsson}
M.~Larsson, B.~Olsson, and P.~Sigray, ``Theoretical study of the co2+
  dication,''
  \href{http://dx.doi.org/https://doi.org/10.1016/0301-0104(89)80157-0}{{\em
  Chemical Physics} {\bfseries 139} no.~2, (1989) 457 -- 469}.
  \url{http://www.sciencedirect.com/science/article/pii/0301010489801570}.

\bibitem{LCederbaum}
G.~Handke, F.~Tarantelli, and L.~S. Cederbaum, ``Triple ionization of carbon
  monoxide,'' \href{http://dx.doi.org/10.1103/PhysRevLett.76.896}{{\em Phys.
  Rev. Lett.} {\bfseries 76} (Feb, 1996) 896--899}.
  \url{https://link.aps.org/doi/10.1103/PhysRevLett.76.896}.

\bibitem{PWThulstrup}
P.~W. Thulstrup, E.~W. Thulstrup, A.~Andersen, and
  Y.~{\ifmmode\ddot{O}\else\"{O}\fi}hrn, ``{Configuration interaction
  calculations of some observed states of NO{-}, NO, NO+, and NO2+},''
  \href{http://dx.doi.org/10.1063/1.1680845}{{\em J. Chem. Phys.} {\bfseries
  60} no.~10, (May, 1974) 3975--3980}.

\bibitem{PhysRevA.100.053422}
T.~Endo, H.~Fujise, H.~Hasegawa, A.~Matsuda, M.~Fushitani, O.~I. Tolstikhin,
  T.~Morishita, and A.~Hishikawa, ``Angle dependence of dissociative tunneling
  ionization of no in asymmetric two-color intense laser fields,''
  \href{http://dx.doi.org/10.1103/PhysRevA.100.053422}{{\em Phys. Rev. A}
  {\bfseries 100} (Nov, 2019) 053422}.
  \url{https://link.aps.org/doi/10.1103/PhysRevA.100.053422}.

\bibitem{Erman_1996}
P.~Erman, P.~A. Hatherly, A.~Karawajczyk, U.~Köble, E.~Rachlew-Källne,
  M.~Stankiewicz, and K.~Y. Franz{\'{e}}n, ``Fragmentation processes of the
  core-excited {NO} molecule,''
  \href{http://dx.doi.org/10.1088/0953-4075/29/8/014}{{\em Journal of Physics
  B: Atomic, Molecular and Optical Physics} {\bfseries 29} no.~8, (Apr, 1996)
  1501--1513}. \url{https://doi.org/10.1088/0953-4075/29/8/014}.

\bibitem{Lai_2016}
W.~Lai and C.~Guo, ``The role of molecular electron distribution in
  strong-field ionization and dissociation of heteronuclear molecules,''
  \href{http://dx.doi.org/10.1088/0953-4075/49/22/225601}{{\em Journal of
  Physics B: Atomic, Molecular and Optical Physics} {\bfseries 49} no.~22,
  (Oct, 2016) 225601}. \url{https://doi.org/10.1088/0953-4075/49/22/225601}.

\bibitem{DLCooper}
D.~L. Cooper, ``{Ab initio investigation of low-lying 2{$\Sigma$}+ and 2{$\Pi$}
  states of NO2+},'' \href{http://dx.doi.org/10.1016/0009-2614(86)80629-7}{{\em
  Chem. Phys. Lett.} {\bfseries 132} no.~4, (Dec, 1986) 377--382}.

\bibitem{PhysRevA.75.062709}
J.~Rajput and C.~P. Safvan, ``Kinetic energy distributions in ion-induced
  $\mathrm{CO}$ fragmentation: Signature of shallow states in multiply charged
  $\mathrm{CO}$,'' \href{http://dx.doi.org/10.1103/PhysRevA.75.062709}{{\em
  Phys. Rev. A} {\bfseries 75} (Jun, 2007) 062709}.
  \url{https://link.aps.org/doi/10.1103/PhysRevA.75.062709}.

\bibitem{Shubhadeep}
S.~Biswas and L.~C. Tribedi, ``A recoil ion momentum spectrometer for probing
  ionization, e-capture, and capture-ionization induced molecular fragmentation
  dynamics,'' \href{http://dx.doi.org/10.1063/5.0068307}{{\em Review of
  Scientific Instruments} {\bfseries 92} no.~12, (2021) 123304},
  \href{http://arxiv.org/abs/https://doi.org/10.1063/5.0068307}{{\ttfamily
  https://doi.org/10.1063/5.0068307}}. \url{https://doi.org/10.1063/5.0068307}.

\bibitem{LEIBF2}
A.~Kumar, J.~Rajput, T.~Sairam, M.~Jana, L.~Nair, and C.~Safvan, ``Setup for
  measuring angular anisotropies in slow ion-molecule collisions,''
  \href{http://dx.doi.org/https://doi.org/10.1016/j.ijms.2014.09.018}{{\em
  International Journal of Mass Spectrometry} {\bfseries 374} (2014) 44--48}.
  \url{https://www.sciencedirect.com/science/article/pii/S1387380614004035}.

\bibitem{MOLPRO}
H.-J. Werner, P.~J. Knowles, G.~Knizia, F.~R. Manby, and M.~Schütz, ``Molpro:
  a general-purpose quantum chemistry program package,''
  \href{http://dx.doi.org/10.1002/wcms.82}{{\em WIREs Computational Molecular
  Science} {\bfseries 2} no.~2, (2012) 242--253},
  \href{http://arxiv.org/abs/https://onlinelibrary.wiley.com/doi/pdf/10.1002/wcms.82}{{\ttfamily
  https://onlinelibrary.wiley.com/doi/pdf/10.1002/wcms.82}}.
  \url{https://onlinelibrary.wiley.com/doi/abs/10.1002/wcms.82}.

\bibitem{doi:10.1063/1.455556}
H.~Werner and P.~J. Knowles, ``An efficient internally contracted
  multiconfiguration–reference configuration interaction method,''
  \href{http://dx.doi.org/10.1063/1.455556}{{\em The Journal of Chemical
  Physics} {\bfseries 89} no.~9, (1988) 5803--5814},
  \href{http://arxiv.org/abs/https://doi.org/10.1063/1.455556}{{\ttfamily
  https://doi.org/10.1063/1.455556}}. \url{https://doi.org/10.1063/1.455556}.

\bibitem{KNOWLES1988514}
P.~J. Knowles and H.-J. Werner, ``An efficient method for the evaluation of
  coupling coefficients in configuration interaction calculations,''
  \href{http://dx.doi.org/https://doi.org/10.1016/0009-2614(88)87412-8}{{\em
  Chemical Physics Letters} {\bfseries 145} no.~6, (1988) 514--522}.
  \url{https://www.sciencedirect.com/science/article/pii/0009261488874128}.

\bibitem{doi:10.1063/1.3609809}
K.~R. Shamasundar, G.~Knizia, and H.-J. Werner, ``A new internally contracted
  multi-reference configuration interaction method,''
  \href{http://dx.doi.org/10.1063/1.3609809}{{\em The Journal of Chemical
  Physics} {\bfseries 135} no.~5, (2011) 054101},
  \href{http://arxiv.org/abs/https://doi.org/10.1063/1.3609809}{{\ttfamily
  https://doi.org/10.1063/1.3609809}}. \url{https://doi.org/10.1063/1.3609809}.

\bibitem{cccbdb}
 \url{http://cccbdb.nist.gov/}.

\bibitem{Huber1979}
K.~P. Huber and G.~Herzberg, {\em Molecular spectra and molecular structure.
  IV. Constants of diatomic molecules}.
\newblock Van Nostrand Reinhold, New York, 1979.

\bibitem{5}
P.~Kumar and N.~P. Sathyamurthy, ``Potential energy curves for neutral and
  multiply charged carbon monoxide,''
  \href{http://dx.doi.org/https://doi.org/10.1007/s12043-010-0006-y}{{\em
  Pramana-J Phys} {\bfseries 74} (Jan, 2010) 49--55}.
  \url{https://link.springer.com/article/10.1007/s12043-010-0006-y#citeas}.

\bibitem{Mathur}
D.~Mathur, E.~Krishnakumar, K.~Nagesha, V.~R. Marathe, V.~Krishnamurthi, F.~A.
  Rajgara, and U.~T. Raheja, ``Dissociation of highly charged co q+ (q>or=2)
  ions via non-coulombic potential energy curves,'' {\em Journal of Physics B:
  Atomic, Molecular and Optical Physics} {\bfseries 26} no.~6, (1993) L141.
  \url{http://stacks.iop.org/0953-4075/26/i=6/a=005}.

\bibitem{WATSON1975384}
W.~Watson, D.~Stewart, A.~Gardner, and M.~Lynch, ``The photoabsorption
  coefficients of co and co2 in the region 350 to 650Å,''
  \href{http://dx.doi.org/https://doi.org/10.1016/0032-0633(75)90144-0}{{\em
  Planetary and Space Science} {\bfseries 23} no.~2, (1975) 384--386}.
  \url{https://www.sciencedirect.com/science/article/pii/0032063375901440}.

\bibitem{NIST_ASD}
A.~Kramida, {Yu.~Ralchenko}, J.~Reader, and {and NIST ASD Team}. {NIST Atomic
  Spectra Database (ver. 5.8), [Online]. Available:
  {\tt{https://physics.nist.gov/asd}} [2021, June 5]. National Institute of
  Standards and Technology, Gaithersburg, MD.}, 2020.

\bibitem{PhysRevA.74.032701}
J.~Rajput, S.~De, A.~Roy, and C.~P. Safvan, ``Kinetic energy distributions and
  signature of target excitation in ${\mathrm{n}}_{2}$ fragmentation on
  collisions with ${\mathrm{ar}}^{9+}$ ions,''
  \href{http://dx.doi.org/10.1103/PhysRevA.74.032701}{{\em Phys. Rev. A}
  {\bfseries 74} (Sep, 2006) 032701}.
  \url{https://link.aps.org/doi/10.1103/PhysRevA.74.032701}.

\bibitem{TMasuoka1}
T.~Masuoka, ``Kinetic energy release in the dissociation of co2+,''
  \href{http://dx.doi.org/10.1063/1.468192}{{\em The Journal of Chemical
  Physics} {\bfseries 101} no.~1, (1994) 322},
  \href{http://arxiv.org/abs/https://doi.org/10.1063/1.468192}{{\ttfamily
  https://doi.org/10.1063/1.468192}}. \url{https://doi.org/10.1063/1.468192}.

\bibitem{Xiaokai}
X.~Li, J.~Yu, H.~Xu, X.~Yu, Y.~Yang, Z.~Wang, P.~Ma, C.~Wang, F.~Guo, Y.~Yang,
  S.~Luo, and D.~Ding, ``Multiorbital and excitation effects on dissociative
  double ionization of co molecules in strong circularly polarized laser
  fields,'' \href{http://dx.doi.org/10.1103/PhysRevA.100.013415}{{\em Phys.
  Rev. A} {\bfseries 100} (Jul, 2019) 013415}.
  \url{https://link.aps.org/doi/10.1103/PhysRevA.100.013415}.

\bibitem{Watson1996Feb}
R.~L. Watson, G.~Sampoll, V.~Horvat, and O.~Heber, ``{Kinetic-energy release in
  the dissociative capture-ionization of CO molecules by 97-MeV
  ${\mathrm{Ar}}^{14+}$ ions},''
  \href{http://dx.doi.org/10.1103/PhysRevA.53.1187}{{\em Phys. Rev. A}
  {\bfseries 53} no.~2, (Feb, 1996) 1187--1190}.

\bibitem{Wells2005Aug}
E.~Wells, V.~Krishnamurthi, K.~D. Carnes, N.~G. Johnson, H.~D. Baxter,
  D.~Moore, K.~M. Bloom, B.~M. Barnes, H.~Tawara, and I.~Ben-Itzhak,
  ``{Proton--carbon monoxide collisions from 10 keV to 14 MeV},''
  \href{http://dx.doi.org/10.1103/PhysRevA.72.022726}{{\em Phys. Rev. A}
  {\bfseries 72} no.~2, (Aug, 2005) 022726}.

\bibitem{Ben-Itzhak1995Jan}
I.~Ben-Itzhak, S.~G. Ginther, V.~Krishnamurthi, and K.~D. Carnes,
  ``{Kinetic-energy release in CO dissociation caused by fast
  ${\mathrm{F}}^{4+}$ impact},''
  \href{http://dx.doi.org/10.1103/PhysRevA.51.391}{{\em Phys. Rev. A}
  {\bfseries 51} no.~1, (Jan, 1995) 391--399}.

\bibitem{Folkerts1996Oct}
H.~O. Folkerts, R.~Hoekstra, and R.~Morgenstern, ``{Velocity and Charge State
  Dependences of Molecular Dissociation Induced by Slow Multicharged Ions},''
  \href{http://dx.doi.org/10.1103/PhysRevLett.77.3339}{{\em Phys. Rev. Lett.}
  {\bfseries 77} no.~16, (Oct, 1996) 3339--3342}.

\bibitem{PhysRevLett.75.1058}
M.~Lundqvist, P.~Baltzer, D.~Edvardsson, L.~Karlsson, and B.~Wannberg, ``Novel
  time of flight instrument for doppler free kinetic energy release
  spectroscopy,'' \href{http://dx.doi.org/10.1103/PhysRevLett.75.1058}{{\em
  Phys. Rev. Lett.} {\bfseries 75} (Aug, 1995) 1058--1061}.
  \url{https://link.aps.org/doi/10.1103/PhysRevLett.75.1058}.

\bibitem{Curtis1984Oct}
J.~M. Curtis and R.~K. Boyd, ``{Predissociation processes in N2+2, O2+2, and
  NO2+ studied by ion kinetic energy spectroscopy},''
  \href{http://dx.doi.org/10.1063/1.448051}{{\em J. Chem. Phys.} {\bfseries 81}
  no.~7, (Oct, 1984) 2991--3001}.

\bibitem{Masuoka1994May}
T.~Masuoka, ``{Kinetic{-}energy release in the dissociation of NO2+},''
  \href{http://dx.doi.org/10.1063/1.467051}{{\em J. Chem. Phys.} {\bfseries
  100} no.~9, (May, 1994) 6422--6428}.

\bibitem{YBKim}
Y.~B. Kim, K.~Stephan, E.~Märk, and T.~D. Märk, ``Single and double
  ionization of nitric oxide by electron impact from threshold up to 180 ev,''
  \href{http://dx.doi.org/10.1063/1.441082}{{\em The Journal of Chemical
  Physics} {\bfseries 74} no.~12, (1981) 6771--6776},
  \href{http://arxiv.org/abs/https://doi.org/10.1063/1.441082}{{\ttfamily
  https://doi.org/10.1063/1.441082}}. \url{https://doi.org/10.1063/1.441082}.

\bibitem{0953-4075-30-24-019}
H.~O. Folkerts, F.~W. Bliek, M.~C. de~Jong, R.~Hoekstra, and R.~Morgenstern,
  ``Dissociation of co induced by $he^{2+}$ ions: I. fragmentation and kinetic
  energy release spectra,'' {\em Journal of Physics B: Atomic, Molecular and
  Optical Physics} {\bfseries 30} no.~24, (1997) 5833.
  \url{http://stacks.iop.org/0953-4075/30/i=24/a=019}.

\bibitem{PhysRevA.47.2827}
I.~Ben-Itzhak, S.~G. Ginther, and K.~D. Carnes, ``Multiple-electron removal and
  molecular fragmentation of co by fast ${\mathit{f}}^{4+}$ impact,''
  \href{http://dx.doi.org/10.1103/PhysRevA.47.2827}{{\em Phys. Rev. A}
  {\bfseries 47} (Apr, 1993) 2827--2837}.
  \url{https://link.aps.org/doi/10.1103/PhysRevA.47.2827}.

\bibitem{PhysRevA.46.3929}
K.~Wohrer, G.~Sampoll, R.~L. Watson, M.~Chabot, O.~Heber, and V.~Horvat,
  ``Dissociation of multicharged co molecular ions produced in collisions with
  97-mev ${\mathrm{ar}}^{14+}$: Dissociation fractions and branching ratios,''
  \href{http://dx.doi.org/10.1103/PhysRevA.46.3929}{{\em Phys. Rev. A}
  {\bfseries 46} (Oct, 1992) 3929--3934}.
  \url{https://link.aps.org/doi/10.1103/PhysRevA.46.3929}.

\bibitem{Suzuki1995}
I.~H. Suzuki and N.~Saito, ``{Ionic Fragmentation of NO Following Excitation of
  the NK-Shell and the OK-Shell Electron},''
  \href{http://dx.doi.org/10.1155/1995/31767}{{\em Laser Chem.} {\bfseries 16}
  no.~1, (1995) 5--18}.

\bibitem{GWei_2009}
G.~Wei, Z.~Jing-Yi, W.~Bing-Xing, W.~Yan-Qiu, and W.~Li, ``Fragmentation of
  {CO} in femtosecond laser fields,''
  \href{http://dx.doi.org/10.1088/0256-307x/26/1/013201}{{\em Chinese Physics
  Letters} {\bfseries 26} no.~1, (Jan, 2009) 013201}.
  \url{https://doi.org/10.1088/0256-307x/26/1/013201}.

\end{thebibliography}\endgroup

\end{document}